\documentclass[
prd,
twocolumn%
,floatfix
,showpacs ,amssymb, aps]{revtex4}
\usepackage{graphicx}
\input{epsf}

\usepackage{amsmath,amssymb}
\usepackage{bm}

\newcommand{\dalm}{\kern1pt\vbox{\hrule height 0.9pt\hbox{\vrule width 0.9pt
\hskip 2.5pt\vbox{\vskip 5.5pt}\hskip 3pt\vrule width 0.3pt}\hrule height 0.3pt}
\kern1pt}

\usepackage{natbib}
\usepackage{wasysym}

\newcommand{\U}{\mathcal{U}}
\newcommand{\g}{\tilde{g}}
\newcommand{\F}{\mathcal{F}}
\newcommand{\T}{\tilde{T}}
\newcommand{\rhot}{\tilde{\rho}}
\newcommand{\Pt}{\tilde{P}}
\newcommand{\ut}{\tilde{u}}

\begin{document}

\title{Structure of Neutron Stars in Tensor-Vector-Scalar Theory}

\author{Paul D. Lasky}
	\email{plasky@astro.swin.edu.au}
	\altaffiliation{Present Address: Centre for Astrophysics and Supercomputing, Swinburne University, Hawthorn VIC 3122, Australia}
	\altaffiliation{School of Mathematical Sciences, Monash University, Wellington Rd., Melbourne 3800, Australia}
\author{Hajime Sotani}
	\email{sotani@astro.auth.gr}
	\affiliation{Theoretical Astrophysics, Eberhard Karls University of T\"ubingen, T\"ubingen 72076, Germany}
\author{Dimitrios Giannios}
	\email{giannios@MPA-Garching.MPG.DE}
	\affiliation{Max Planck Institute for Astrophysics, Box 1317, D-85741 Garching, Germany}

		\begin{abstract}
		Bekenstein's Tensor-Vector-Scalar (TeVeS) theory has had considerable success in explaining various phenomena without the need for dark matter.  However, it is difficult to observationally discern the differences between TeVeS and predictions made within the $\Lambda$-cold dark matter concordance model.  This implies that alternative tests are required that independently verify which theory is correct.  For this we turn to the strong-field regime of TeVeS.  In particular, we solve the spherically symmetric equations of hydrostatic equilibrium for a perfect fluid with a realistic equation of state to build models of neutron stars in TeVeS.  We show that causality within the neutron star is only maintained for certain cosmological values of the scalar field, which allows us to put constraints on this value independently of cosmological observations.  We also discuss in detail the internal structure of neutron stars and how each of the free parameters in the theory effects the overall size and mass of the neutron stars.  In particular, the radii of neutron stars in TeVeS can significantly differ from those in General Relativity for certain values of the vector field coupling, which allows us to also place extra constraints on this parameter.  Finally, we discuss future observations of neutron stars using both the electromagnetic and gravitational wave spectrums that will allow for tests of the appropriate theory of gravity.
		\end{abstract}
		
		\pacs{04.50.Kd, 97.60.Jd, 04.80.Cc}

		\maketitle

\section{Introduction}
Modified Newtonian Dynamics (MOND) \cite{milgrom83} was developed to describe the differences between the observed masses of galaxies and clusters of galaxies with the masses inferred from Newtonian dynamics.  To this task, MOND has been extremely successful (for a review see \cite{sanders02} and references therein), however it is clear that MOND was never more than a toy model for gravitation as it is not a covariant theory.  A relativistic version has been put forward by \citet{bekenstein04}, where the standard Einstein tensor field of General Relativity (GR) is coupled to a vector field as well as a scalar field, hence the theory is called Tensor-Vector-Scalar (TeVeS).

In the weak acceleration limit, TeVeS reproduces MOND, and hence carries forth all the successes of that theory.  In the Newtonian limit, the parametrized post-Newtonian (PPN) coefficients, $\beta$ and $\gamma$, agree with the results of solar system tests.  Moreover, on cosmological scales the theory has been tested against a host of gravitational lensing data \cite{chen06,zhao06}, and can also be shown to reproduce key features of the cosmic microwave background radiation \cite{skordis06} as well as galaxy distributions \cite{dodelson06}.  Whilst the reproduction of known observations is obviously critical to the success of any theory, one also desires predictions that differ from the standard theory, which here is the $\Lambda$CDM concordance model.  To this we turn to the strong-field limit in order to study both black holes and neutron stars.

Black holes in TeVeS were first studied by \citet{giannios05}, who solved the field equations for static, spherically symmetric spacetimes in vacuum, hence deriving the Schwarzschild-TeVeS solution.  He found two distinct branches of solutions dependant on the form of the vector field.  When the vector field is aligned with only the temporal direction, Giannios was able to show that the PPN coefficients were identical to those of a Schwarzschild black hole in GR.  Therefore, in this regime the predictions of TeVeS are observationally identical to GR.  The second class of solutions studied by Giannios also included a non-zero radial component to the vector field.  Here, he was able to show that the PPN parameters differ from those predicted by GR.  \citet{sagi08} then extended this line of work and showed the solution of a static, spherically symmetric system with the inclusion of electromagnetic fields leads identically to the Reissner-Nordstr\"om spacetime when the vector field has a vanishing radial component.  Moreover, Sagi \& Bekenstein further showed that black hole thermodynamics of these spacetimes are identical to those in GR.  To date, to the best of our knowledge, the equations of hydrostatic equilibrium in TeVeS have not been presented, and therefore there is no consistent study of compact stars.

The study of compact stars in alternative theories of gravity is of critical importance.  Indeed in many theories such as $f(R)$ gravities and scalar-tensor theories, one can derive observable deviations from GR that can be predicted with electromagnetic observations (for a review see \cite{psaltis08}).  It is therefore crucial to the prosperity of Bekenstein's TeVeS that compact stars are studied such that definitive observational predictions can be made that differentiate the theory from its counterparts.

In this paper we derive the TeVeS Tolman-Oppenheimer-Volkoff (TOV) equations of hydrostatic equilibrium for a spherically symmetric perfect fluid.  As a first approximation we consider spacetimes where the vector field has only a non-vanishing temporal component.  The equations are expressed in such a way that the deviations from both GR and scalar-tensor theories are obvious.  Applying a realistic polytropic equation of state allows us to numerically solve the system of equations, thereby deriving realistic neutron star models.  In the strong-field limit, TeVeS has three free parameters, which are the strengths of the vector and scalar field couplings, $K$ and $k$ respectively, as well as the cosmological value of the scalar field $\varphi_{c}$.  We argue that the requirement of causality can place restrictions on the scalar field such that $\varphi_{c}\gtrsim0.001$.  Moreover, we use current observations of neutron stars to constrain $K\lesssim1$.

The paper is structured as follows: In section \ref{eqs} we provide a brief description of the main TeVeS field equations.  We do not provide a detailed derivation of these equations, but instead point the interested reader to ref. \cite{bekenstein04}.  In section \ref{TOV} we derive the TOV equations of hydrostatic equilibrium and further discuss limits of the equations as the theory tends towards scalar-tensor theories and GR.  In section \ref{NSM} we construct neutron star models, in particular discussing the equation of state in section \ref{EOS} before going on to vary the cosmological value of the scalar field, vector and scalar coupling parameters in sections \ref{CVSF}, \ref{VFC} and \ref{SFC} respectively.  In section \ref{Obs} we discuss the observational consequences of the TeVeS models as distinct from GR, and finally conclude in section \ref{Conc}.  Unless otherwise specified, coordinates used are $x^{\mu}=\left(t,r,\theta,\phi\right)$, where Greek indices run over $0\ldots3$.  Antisymmetrization is denoted by square brackets, i.e. $A_{[\mu\nu]}:=A_{\mu\nu}-A_{\nu\mu}$, and round brackets denote symmetrization, $A_{(\mu\nu)}:=A_{\mu\nu}+A_{\mu\nu}$.

\section{TeVeS Equations}\label{eqs}
\subsection{Field Equations}\label{fieldeqs}
TeVeS is based on three gravitational fields; an Einstein metric, $g_{\mu\nu}$, which has a well defined inverse, $g^{\mu\nu}$, a timelike vector field, $\U_{\mu}$, and a dynamical scalar field $\varphi$.  The physical metric, $\g_{\mu\nu}$, is defined in terms of the Einstein metric, the vector and scalar fields according to
\begin{align}
	\g_{\mu\nu}=e^{-2\varphi}\left(g_{\mu\nu}+\U_{\mu}\U_{\nu}\right)-e^{2\varphi}\U_{\mu}\U_{\nu}.\label{gtilde}
\end{align}
The field equations are then derived by varying the total action of the theory (for details see ref. \cite{bekenstein04}) with respect to the Einstein tensor, the vector field, the dynamical scalar field and also a non-dynamical scalar field, $\sigma$.  By varying the action with respect to both scalar fields, $\sigma$ and $\varphi$, one can show
\begin{align}
	&\left[\mu\left(k\ell^{2} h^{\alpha\beta}\varphi_{,\alpha}\varphi_{,\beta}\right)h^{\gamma\delta}\varphi_{,\gamma}\right]_{;\delta}\notag\\
		&\qquad=kG\left[g^{\alpha\beta}+\left(1+e^{-4\varphi}\right)\U^{\alpha}\U^{\beta}\right]\T_{\alpha\beta},\label{scalar}
\end{align}
where a semi-colon denotes covariant differentiation with respect to the Einstein tensor and $h_{\mu\nu}:=g_{\mu\nu}-\U_{\mu}\U_{\nu}$.  A positive dimensionless scalar, $k$, and another positive scalar of dimension length, $\ell$, have been introduced that act to couple the scalar fields to gravity.  Moreover, $\mu(x)$ is defined with respect to a free-function $\F(\mu)$, which is not predicted by the theory.  The relation between these two functions is given by the following differential equation;
\begin{align}
	-\mu \F(\mu)-\frac{1}{2}\mu^{2}\frac{d\F}{d\mu}=x.
\end{align}
We note here that \citet{bekenstein04} has assumed a specific functional form for $\F$, which led him to derive $k=0.03$ using Newtonian limits of TeVeS.  As we will show, the results in this paper are completely independent of the functional form of $\F$.  As such, in section \ref{SFC} we will explore the parameter space around $k$ in an attempt to provide a model independent constraint on its value.

Defining $\sigma$ as a ``non-dynamic'' field implies it has no kinetic term in its equation of motion.  Therefore, the field equation on $\sigma$ is simply an algebraic relation;
\begin{align}
	kG\sigma^{2}=\mu\left(k\ell^{2}h^{\alpha\beta}\varphi_{,\alpha}\varphi_{,\beta}\right).
\end{align}
In this paper we only consider the strong-field limit, which has $\mu=1$ as an excellent approximation (see the detailed discussions in \cite{sagi08,giannios05,bekenstein04}).  Therefore, we are considering a regime where
\begin{align}
	\sigma=\frac{1}{kG}.
\end{align}
Moreover, in this limit one finds that the free function $\F$ does not contribute to the tensor field equations (\ref{Einstein}), and therefore it is ignored from the remainder of this discussion.

In TeVeS, the timelike vector field is normalized, which is given by 
\begin{align}
	g^{\alpha\beta}\U_{\alpha}\U_{\beta}=-1.\label{norm}
\end{align}
Note that the vector field is raised and lowered with respect to the Einstein metric.  The coupling of the vector field to the other fields is given by the positive, dimensionless constant $K$.  One can show that taking the limit as $K$ and $k$ go to zero implies the theory reduces to GR.  Varying the total action with respect to the vector field yields
\begin{align}
	&\,\,K\left({\U^{[\mu;\alpha]}}_{;\alpha}+\U^{\mu}\U_{\alpha}{\U^{[\alpha;\beta]}}_{;\beta}\right)\notag\\
		&+8\pi G\sigma^{2}\left[\U^{\alpha}\varphi_{,\alpha}g^{\mu\beta}\varphi_{,\beta}+\U^{\mu}\left(\U^{\alpha}\varphi_{,\alpha}\right)^{2}\right]\notag\\
		=&8\pi G\left(1-e^{-4\varphi}\right)\left(g^{\mu\alpha}\U^{\beta}\T_{\alpha\beta}+\U^{\mu}\U^{\alpha}\U^{\beta}\T_{\alpha\beta}\right),\label{vec}
\end{align}
where the normalization criteria (\ref{norm}) has been implicitly used in the derivation.  The use of this normalization implies that contraction of equation (\ref{vec}) with respect to the vector field gives zero identically.  This implies that equation (\ref{vec}) is spatial, in the sense that it is orthogonal to the timelike vector field, $\U_{\mu}$.  Moreover, it implies that there are just three components of equation (\ref{vec}), where the fourth component is given by the normalization condition.

We note here that despite the extensive literature surrounding the theory of TeVeS, there is surprisingly little information regarding constraints on the vector coupling parameter, $K$.  To the best of our knowledge, the only discussion is from \citet{sanders06}, who has phenomenologically argued for $K<10^{-7}$ based on solar system experiments, although he has not gone into detail with the calculation.  If $K$ is indeed this small, then the differences in neutron stars in TeVeS from those predicted in GR is not significant enough to be detectable.  We further note here that strange things occur when $K\ge2$ (see section \ref{Kk}), and we therefore tend to keep the rest of our discussions below this value.

Finally, variation of the total action with respect to the Einstein metric yields the modified Einstein equations
\begin{align}
	G_{\mu\nu}=8\pi G\left[\T_{\mu\nu}+\left(1-e^{-4\varphi}\right)\U^{\alpha}\T_{\alpha(\mu}\U_{\nu)}+\tau_{\mu\nu}\right]+\Theta_{\mu\nu},\label{Einstein}
\end{align}
where $\T_{\mu\nu}$ is the physical energy-momentum tensor field which is defined using the physical metric and
\begin{align}
	\tau_{\mu\nu}:=&\sigma^{2}\Big[\varphi_{,\mu}\varphi_{,\nu}-\frac{1}{2}g^{\alpha\beta}\varphi_{,\alpha}\varphi_{,\beta}g_{\mu\nu}-\frac{G\sigma^{2}}{4\ell^{2}}F(kG\sigma^{2})g_{\mu\nu}\notag\\
		&-\U^{\alpha}\varphi_{,\alpha}\left(\U_{(\mu}\varphi_{,\nu)}-\frac{1}{2}\U^{\beta}\varphi_{,\beta}g_{\mu\nu}\right)\Big],\label{tau}\\
	\Theta_{\mu\nu}:=&K\Big(g^{\alpha\beta}\U_{[\alpha,\mu]}\U_{[\beta,\nu]}-\frac{1}{4}g^{\gamma\delta}g^{\alpha\beta}\U_{[\gamma,\alpha]}\U_{[\delta,\beta]}g_{\mu\nu}\notag\\
		&-\U_{\mu}\U_{\nu}\U_{\alpha}{\U^{[\alpha;\beta]}}_{;\beta}\Big)-8\pi G\U_{\mu}\U_{\nu}\Big[\sigma^{2}\left(\U^{\alpha}\varphi_{,\alpha}\right)^{2}\notag\\
		&-\left(1-e^{-4\varphi}\right)\T_{\alpha\beta}\U^{\alpha}\U^{\beta}\Big].\label{Theta}
\end{align}

\subsection{Fluid equations}
We wish to consider here a perfect fluid, which is described in the physical frame.  Therefore, the energy momentum tensor is
\begin{align}
	\T_{\mu\nu}=\left(\rhot+\Pt\right)\ut_{\mu}\ut_{\nu}+\Pt \g_{\mu\nu},
\end{align}
where $\ut_{\mu}$, $\rhot$ and $\Pt$ are the fluid four-velocity, energy-density and isotropic pressure measured in the physical frame.  In the next section we will choose the fluid four velocity and the timelike vector field to be aligned.  In this case, \citet{bekenstein04} showed $\ut_{\mu}=e^{\varphi}\U_{\mu}$, which further implies
\begin{align}
	\T_{\mu\nu}=e^{2\varphi}\rhot\U_{\mu}\U_{\nu}+e^{-2\varphi}\Pt\left(g_{\mu\nu}+\U_{\mu}\U_{\nu}\right).
\end{align}

Finally, conservation of energy-momentum is given by
\begin{align}
	\T^{\alpha}{}_{\mu |\alpha}=0,\label{cons}
\end{align}
where a vertical bar represents covariant differentiation with respect to the physical metric (i.e. $\g_{\mu\nu |\sigma}=0$) and $\T^{\mu}{}_{\nu}:=\g^{\mu\alpha}\T_{\alpha\nu}$.

\section{Hydrostatic Equilibrium}\label{TOV}
The aim of this paper is to derive the equations of hydrostatic equilibrium that are solutions of the TeVeS field equations outlined in the previous section.  For this we shall assume a static, spherically symmetric form for the Einstein metric \footnote{We note that, due to our subsequent choice of the vector field, choosing the physical metric rather than the Einstein metric would give the same results after scaling the coordinates with an appropriate factor of the scalar field.}.  Moreover, for this paper we shall choose to work in Schwarzschild coordinates, as opposed to isotropic coordinates which seem to be the popular choice in TeVeS.  We do this so that the equations reduce to the familiar form of the TOV equations in the general relativistic limit.  The metric is 
\begin{align}
	ds^{2}=-e^{\nu}dt^{2}+e^{\zeta}dr^{2}+r^{2}d\Omega^{2},\label{Emetric}
\end{align}
where $ds^{2}:=g_{\alpha\beta}dx^{\alpha}dx^{\beta}$ is the Einstein line element and $d\Omega^{2}:=d\theta^{2}+\sin^{2}\theta d\phi^{2}$ is the line element for the two spheres.  Moreover, the spacetime being static and spherically symmetric implies all functions depend only on the radial coordinate.

Analogously to choosing an ansatz for the tensor field, we must also select an ansatz for the vector field.  In the cosmological limit the vector field must only have a timelike component such that the theory does not pick out a preferred direction in space.  However, near strong gravitational fields this is not required.  Instead, the only requirement is that the asymptotic limit of any spatial components must vanish.  Having said that, out of mathematical convenience we choose in this article to only consider a vector field where the spatial components vanish.  It is noted that this is restricting our solution space, and a future direction for this research is to reintroduce a non-zero radial component into the vector field. 

With this choice of vector field, the normalization criteria (\ref{norm}), together with the Einstein metric (\ref{Emetric}) uniquely determines the form of the vector field to be
\begin{align}
	\U^{\mu}=\left[e^{-\nu/2},\,0,\,0,\,0\right].
\end{align}
In section \ref{fieldeqs} we showed that the vector field equation (\ref{vec}) only provides three components of the vector field, and the fourth is given by the normalization criteria.  This implies that, given the form of the vector field above, equation (\ref{vec}) is trivially satisfied, and indeed a simple calculation reveals this to be true.

Utilizing the form of the vector and tensor fields along with the definition of the physical metric (\ref{gtilde}), it is trivial to show that the physical line element is
\begin{align}
	d\tilde{s}^{2}=-e^{\nu+2\varphi}dt^{2}+e^{\zeta-2\varphi}dr^{2}+e^{-2\varphi}r^{2}d\Omega^{2},\label{Pmetric}
\end{align}
where $d\tilde{s}^{2}:=\g_{\alpha\beta}dx^{\alpha}dx^{\beta}$.

The conservation equations can now be expressed in terms of the metric coefficients.  Only the radial component of equation (\ref{cons}) is non-trivially satisfied, which is Euler's equation describing the pressure gradient as a function of the metric coefficients
\begin{align}
	-\Pt'=\left(\frac{\nu'}{2}+\varphi'\right)\left(\rhot+\Pt\right),\label{Euler}
\end{align}
where a prime denotes differentiation with respect to the radial coordinate.

The scalar field equaiton (\ref{scalar}) can be shown to reduce to 
\begin{align}
	\frac{e^{-\left(\nu+\zeta\right)/2}}{r^{2}}\left[\varphi'r^{2}e^{\left(\nu-\zeta\right)/2}\right]'=kGe^{-2\varphi}\left(\rhot+3\Pt\right).\label{sc1}
\end{align}
Integration of equation (\ref{sc1}) yields
\begin{align}
	\varphi'=\frac{kGM_{\varphi}e^{\left(\zeta-\nu\right)/2}}{4\pi r^{2}},\label{sc2}
\end{align}
where the ``scalar mass'' $M_{\varphi}$, has been defined according to \cite{bekenstein04}
\begin{align}
	M_{\varphi}(r):=4\pi\int_{0}^{r}\left(\rhot+3\Pt\right)e^{\left(\nu+\zeta\right)/2-2\varphi}r^{2}dr.\label{scalarmass}
\end{align}

The only remaining field equations are the modified Einstein equations given by (\ref{Einstein}).  For completeness, the individual components of the Einstein equations are given in appendix \ref{app}.  After much algebra one can show the modified Einstein equations reduce to 
\begin{align}
	r\zeta'+e^{\zeta}-1=&8\pi G\rhot e^{\zeta-2\varphi}r^{2}+\frac{4\pi r^{2}}{k}\varphi'^{2}\notag\\
		&+\frac{Kr^{2}}{4}\left(\frac{\nu'^{2}}{2}-\nu'\zeta'+2\nu''+\frac{4\nu'}{r}\right),\label{Eins1}\\
	r\nu'-e^{\zeta}+1=&8\pi Ge^{\zeta-2\varphi}\Pt r^{2}+\frac{4\pi r^{2}}{k}\varphi'^{2}-\frac{Kr^{2}}{8}\nu'^{2},\label{Eins2}\\
	\frac{r}{4}\big(2\nu'-2\zeta'-&\nu'\zeta'r+2\nu''r+\nu'^{2}r\big)\notag\\
	=&8\pi G\Pt e^{\zeta-2\varphi}r^{2}-\frac{4\pi r^{2}}{k}\varphi'^{2}+\frac{Kr^{2}}{8}\nu'^{2}.\label{Eins3}
\end{align}
We note that in GR (which is recovered by taking the limit $K=k=0$), equation (\ref{Eins1}) is both independent of the metric coefficient, $\nu$, and is also first-order in all derivatives.  In TeVeS however, the dependance of equation (\ref{Eins1}) on $\nu$, and also the second-order terms are due to the presence of the vector field.  These extra terms complicate the equations considerably, as will be seen in more detail in the following section.

\subsection{TOV equations}
We desire a form of the above equations that resemble the familiar TOV equations of hydrostatic equilibrium of GR.  To this end we define a new function, $m(r)$, according to 
\begin{align}
	e^{-\zeta}=1-\frac{2m}{r}.\label{ezeta}
\end{align}

By adding equations (\ref{Eins2}) and (\ref{Eins3}), and substituting the result through equation (\ref{Eins1}) to eliminate the second-order derivatives in $\nu$, one finds
\begin{align}
	m'&\left(1-\frac{K}{2}\right)-\frac{Km}{2r}=4\pi Ge^{-2\varphi}r^{2}\left(\rhot+2K\Pt\right)\notag\\
		&+\left[\frac{2\pi r^{2}}{k}\varphi'^{2}-\frac{Kr\nu'}{4}\left(1+\frac{r\nu'}{4}\right)\right]\left(1-\frac{2m}{r}\right).\label{TOV1}
\end{align}
Moreover, substituting equation (\ref{ezeta}) into equation (\ref{Eins2}) yields
\begin{align}
	\nu'\left(1+\frac{Kr\nu'}{8}\right)=\frac{8\pi Ge^{-2\varphi}\Pt r^{2}+2m/r}{r\left(1-2m/r\right)}+\frac{4\pi r}{k}\varphi'^{2}.\label{TOV2}
\end{align}

Equations (\ref{Euler}), (\ref{TOV1}) and (\ref{TOV2}), together with the equations describing the scalar field, (\ref{sc2}) and (\ref{scalarmass}), now form the coupled system of equations describing hydrostatic equilibrium in TeVeS.  We will show in the next section that these reduce to the standard TOV equations of both scalar-tensor theories and GR in the appropriate limits.

Equation (\ref{TOV2}) is expressed in an illustrative form such that the differences between GR and TeVeS are obvious (see the discussion in section \ref{limits}).  However, we note that equation (\ref{TOV2}) is not expressed in a convenient form due to the quadratic term in $\nu'$.  We can therefore algebraically solve this equation for $\nu'$, such that it is expressed explicitly;
\begin{align}
	\frac{Kr}{4}\nu'=&-1\notag\\
		&\pm\left\{1+K\left[\frac{4\pi Ge^{-2\varphi}\Pt r^{2}+m/r}{r\left(1-2m/r\right)}+\frac{2\pi r^{2}}{k}\varphi'^{2}\right]\right\}^{1/2}.\label{TOV2a}
\end{align}
One can see that there are two solutions to the above equation, however we note that the negative sign is not physical.  In this case it is trivial to see that the limit $K\rightarrow0$ does not exist.  We therefore exclude this solution from the remainder of the discussion.  


Before moving on, there exists another convenient equation that will subsequently be used that is derived from substituting equations (\ref{TOV2}) through (\ref{TOV1}) to yield
\begin{align}
	\Bigg[m'+\frac{m}{r}-\frac{\nu' r}{2}\left(1-\frac{2m}{r}\right)&+16\pi Ge^{-2\varphi}\Pt r^{2}\Bigg]\left(1-\frac{K}{2}\right)\notag\\
		=4\pi& Ge^{-2\varphi}r^{2}\left(\rhot+3\Pt\right).\label{conv}
\end{align}
From the grouping of the terms on the left hand side of this equation, it can be used to learn much about the range of the vector field coupling constant, $K$.  We analyse this in more detail in section \ref{Kk}.

\subsection{Limits of equations}\label{limits}
\subsubsection{Scalar-Tensor theory}
TeVeS couples the vector field to the other fields through the positive constant $K$.  Setting this to zero is equivalent to ignoring the effects of the vector field in the equations.  This is best seen through the definition of the vector field action, which is directly proportional to $K$ (see eqn. (26) of \citet{bekenstein04}).

With $K=0$, the quadratic term in $\nu'$ in equation (\ref{TOV2}) vanishes.  Moreover, equation (\ref{TOV1}) loses the explicit coupling between the function $m$ and the metric coefficient $\nu$;  
\begin{align} 
	m'&=4\pi G\rhot e^{-2\varphi}r^{2}+\frac{2\pi r^{2}}{k}\varphi'^{2}\left(1-\frac{2m}{r}\right),\\
	\nu'&=\frac{8\pi Ge^{-2\varphi}\Pt r^{2}+2m/r}{r\left(1-2m/r\right)}+\frac{4\pi r}{k}\varphi'^{2}.
\end{align}
These equations are now the same as the well known TOV equations for scalar-tensor gravity \cite{damour93,harada98}.  

\subsubsection{General Relativity}
The GR limit is reached by also letting $k=0$.  We note from equation (\ref{sc2}) that $\varphi'$ is proportional to $k^{2}$, implying $\varphi'/k\propto k$.  In this limit, equation (\ref{sc2}) can be integrated to give $\varphi=\varphi_{c}$ which is constant.  Then
\begin{align}
	m=&4\pi Ge^{-2\varphi_{c}}\int_{0}^{r}\rhot r^{2}dr,\label{hghgh}\\
	\nu'=&\frac{8\pi Ge^{-2\varphi_{c}}\Pt r^{2}+2m/r}{r\left(1-2m/r\right)}\\
		=&\frac{-2\Pt'}{\rhot+\Pt}
\end{align}
which are the standard TOV equations of GR with a constant, background scalar field, $\varphi_{c}$.  One can always choose an appropriate coordinate transformation such that this scalar field vanishes, implying the equations reduce to the standard TOV equations in GR.  Moreover, equation (\ref{hghgh}) is a usual definition for mass, and we therefore refer to the function $m$ as a mass function, even in the TeVeS regime where it is not clear whether this truly represents any physical mass of the system.  

\section{Neutron Star Model}\label{NSM}

In this section we concentrate on realistic neutron star models based on the full set of TOV equations in the TeVeS theory derived in the previous section.  We begin by briefly looking at the spacetime external to the neutron star in order to derive observable quantities, such as the ADM mass of the neutron star.

\subsection{Exterior region}\label{ER}
The boundary of a neutron star, denoted $r_{s}$, is defined as the point at which the pressure vanishes, $\Pt=0$.  For any $r\ge r_{s}$ (for any reasonable equation of state; see below), this implies that the density is also zero, and the spacetime is therefore given by the Schwarzschild-TeVeS solution.  We note that an analytic solution with $\Pt=\rhot=0$ in Schwarzschild coordinates is possible, however its implicit nature provides no new information to the present discussion, and hence we omit its inclusion.  An analytic vacuum solution in a nicer form is known in isotropic coordinates, and was first derived by \citet{giannios05}.  For this discussion we simply require the asymptotic behaviour of the vacuum equations far from the source, such that we can gain an understanding of the function $m$ and the scalar field in the asymptotic limit, and also derive the ADM mass of the system.

For deriving observable quantities in TeVeS we wish to use the physical metric.  However, the physical metric, as expressed in equation (\ref{Pmetric}), is not in an asymptotically flat form.  Therefore, to derive the ADM mass for the system being considered herein, we require a coordinate transformation that brings equation (\ref{Pmetric}) into such a form.  Bearing in mind that the cosmological value of the scalar field is defined as the limit of the scalar field as $r\rightarrow\infty$, we perform a coordinate transformation given by $\hat{t}=te^{\varphi_{c}}$ and $\hat{r}=re^{-\varphi_{c}}$, which yields
\begin{align}
	d\tilde{s}^{2}=-e^{\nu+2\left(\varphi-\varphi_{c}\right)}d\hat{t}^{2}+e^{-2\left(\varphi-\varphi_{c}\right)}\left(e^{\zeta}d\hat{r}^{2}+\hat{r}^{2}d\Omega^{2}\right).\label{Pmetricasymp}
\end{align}

Setting $\rhot=\Pt=0$ in the TOV equations gives the Schwarzschild-TeVeS solution in asymptotically flat coordinates.  By expanding $m$ and $\nu$ in infinite series', one can integrate equation (\ref{sc2}) (noting that $\rhot=\Pt=0$ implies $M_{\varphi}$ is constant) which gives
\begin{align}
	\varphi&=\varphi_{c}-\frac{kGM_{\varphi}}{4\pi e^{\varphi_{c}} \hat{r}}+\mathcal{O}\left(\frac{1}{\hat{r}^{2}}\right),
\end{align}
which is in agreement with \cite{giannios05}.  Further substituting the series expansions through the remaining equations implies the asymptotic form of the Schwarzshild-TeVeS solution can be shown to be
\begin{align}
	\tilde{g}_{\hat{t}\hat{t}}&=-1+\frac{2e^{-\varphi_{c}}}{\hat{r}}\left(m+\frac{kGM_{\varphi}}{4\pi}\right)+\mathcal{O}\left(\frac{1}{\hat{r}^{2}}\right),\\
	\tilde{g}_{\hat{r}\hat{r}}&=1+\frac{2e^{-\varphi_{c}}}{\hat{r}}\left(m+\frac{kGM_{\varphi}}{4\pi}\right)+\mathcal{O}\left(\frac{1}{\hat{r}^{2}}\right).
\end{align}
Therefore, the ADM mass of the system is \cite{misner73}
\begin{align}
	M_{\rm{ADM}}=\left(m_{\infty}+\frac{kGM_{\varphi}}{4\pi}\right)e^{-\varphi_{c}},
\end{align}
where $m_{\infty}$ is the mass function evaluated at radial infinity.  As expected, this implies that the scalar field mass contributes to the entire mass of the spacetime.  In GR, one finds that the mass function outside the star is a constant, and therefore the ADM mass equals the mass function for all $r\ge r_{s}$.  However, in the TeVeS case, the non-zero contribution of the scalar field implies that the mass function still varies outside the star (see figure \ref{massdep}), and that this is not equivalent to the ADM mass of the spacetime.

\subsection{The coupling parameters}\label{Kk}
We can utilize the equations derived herein to put significant constraints on the parameter $K$.  Firstly, by looking at equation (\ref{conv}), one can see that the structure of the equations significantly changes either side of $K=2$.  Indeed, for $K=2$ it can be shown that an equation of state is forced upon the system, namely $\rhot=-3\Pt$.  Moreover, analysing the remaining equations shows that the system has become underdetermined due to a redundancy in the equations.  

We can analytically determine that values of $K>2$ are also unphysical.  This can be seen by performing a series expansion of the hydrostatic equilibrium equations near $r=0$.  After much work, one can show
\begin{align}
	\varphi=&\varphi_{0}+\frac{\varphi_{2}r^{2}}{2}+\mathcal{O}\left(r^{4}\right),\label{phiin}\\
	\nu=&\nu_{0}+\frac{\nu_{2}r^{2}}{2}+\mathcal{O}\left(r^{4}\right),\\
	\Pt=&\Pt_{0}+\frac{\Pt_{2} r^{2}}{2}+\mathcal{O}\left(r^{4}\right),\\
	m=&m_{3}r^{3}+\mathcal{O}\left(r^{5}\right),
\end{align}
where the expansion coefficients are given by
\begin{align}
	\varphi_{2}=&\frac{kG}{3}e^{-2\varphi_{0}}\left(\rhot_{0}+3\Pt_{0}\right),\\
	\nu_{2}=&\frac{16\pi G}{3\left(2-K\right)}e^{-2\varphi_{0}}\left(\rhot_{0}+3\Pt_{0}\right),\\
	\Pt_{2}=&-\frac{\rhot_{0}+\Pt_{0}}{2}\left(2\varphi_{2}+\nu_{2}\right),\\
	m_{3}=&\frac{4\pi G}{3\left(2-K\right)}e^{-2\varphi_{0}}\left(2\rhot_{0}+3K\Pt_{0}\right).\label{m3}
\end{align}
It is trivial to see that $K>2$, and any reasonable equation of state (i.e. such that $\rhot+3\Pt>0$), implies $m_{3}<0$, and hence $m<0$ for small radii.  It is not entirely clear whether a negative mass function itself is unphysical.  However,  it can be shown that for models with $K>2$, the pressure diverges from the stellar centre outward.  This implies a stellar model is not possible as the pressure never vanishes, and hence the surface of the star is not defined.  Therefore, we can restrict the values of the vector coupling parameter to $0<K<2$.  This is consistent with \citet{sagi08}, who showed that physical black hole solutions are only valid for $K<2$.

\subsection{Equation of State}\label{EOS}
As mentioned, the TOV system of equations are closed with an equation of state relating the pressure and the density.  There has been considerable interest over recent years on the neutron star equation of state (see \cite{lattimer07} and references therein).  If GR is the correct theory of gravity, then information can be gained on the appropriate equation of state from the masses and radii of neutron stars.  However, when one considers alternative theories of gravity there is a degeneracy between the equation of state and the free parameters of the theory.  Indeed this is true in TeVeS as we will show below.  

In order to concentrate our attention on the structural dependence of neutron stars on the free-parameters in TeVeS, we choose to only work with a single equation of state in this paper.  It is noted however that this implies we can not stringently constrain these parameters due to the intrinsic uncertainties associated with our choice of equation.  We use a polytropic equation of state given by
\begin{align}
	\Pt=&\mathcal{K}n_{0}m_{b}\left(\frac{\tilde{n}}{m_{b}}\right)^{\Gamma},\\
	\rhot=&\tilde{n}m_{b}+\frac{\Pt}{\Gamma-1},\\
	m_{b}=&1.66\times10^{-24}\,\,{\rm g},\\
	n_{0}=&0.1\,\,{\rm fm}^{-3},
\end{align} 
where $\Gamma=2.46$ and $\mathcal{K}=0.00936$.  Here, the parameters $\Gamma$ and $\mathcal{K}$ have been fitted to the tabulated data for the realistic equation of state, known as equation of state A given in \citet{arnett77}.  This equation of state has been extensively used in previous studies of neutron stars in GR and scalar-tensor theories (see for example \citet{damour93,sotani04}).

The interior conditions for the neutron stars are found using equations (\ref{phiin}-\ref{m3}).  Using these conditions we can determine stellar models for various parameters associated with the theory by numerically solving the TeVeS-TOV equations derived in the previous sections with the aforementioned equation of state.  Numerically, we consider radial infinity to be $r=100\,{\rm km}$, and have verified that the results do not change when we expand our numerical box to $r=500\,{\rm km}$.  The structure of neutron stars depends on the various parameters in the TeVeS theory, for example the vector and scalar field coupling parameters, $K$ and $k$ respectively, and also the cosmological value of the scalar field, $\varphi_{c}$.  We therefore spend a section on the variation of each of these parameters.

\subsection{Cosmological Value of the Scalar Field}\label{CVSF}
Cosmological considerations imply the asymptotic value of the scalar field is restricted to $0\le\varphi_{c}\ll1$ \cite{bekenstein04}.  However, using neutron stars we can place a further restriction on the lower bound for this value, as shown presently.

\begin{figure}[h]
		\begin{center}
		\includegraphics[height=0.3\textwidth,width=0.4\textwidth]{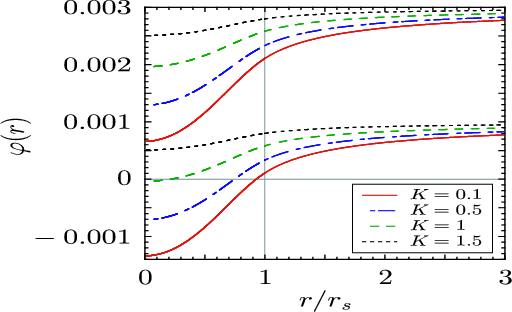}
		\caption{\label{varphi1} Scalar field as a function of radius with fixed $k=0.03$.  Here, the top four lines have $\varphi_c=0.003$ and the bottom four lines have $\varphi_{c}=0.001$.  We exclude all models where $\varphi<0$ for any value of $r$ as this implies violation of causality.}  
		\end{center}
\end{figure}

Bekenstein showed that TeVeS allows for superluminal propagation of tensor, vector and scalar perturbations when $\varphi<0$ (see section VIII of \cite{bekenstein04}).  This was shown using perturbations of the various fields in the physical frame, and was independent of the matter content of the model.  Therefore, for neutron star models we demand that the scalar field be everywhere greater than or equal to zero.  This can be used to constrain the cosmological value of the scalar field.  Figure \ref{varphi1} shows the scalar field as a function of the radius for various values of the vector coupling parameter.  Here, each line corresponds to the stellar model with the maximum mass (all figures henceforth that are functions of the radius also correspond to the models with the maximum mass).  In fig. \ref{varphi1}, $k=0.03$, the top series of lines have $\varphi_{c}=0.003$ and the bottom are for $\varphi_{c}=0.001$.  One can see that for $\varphi_{c}=0.001$, models where $K\lesssim1$ are unphysical as the interior of the star has $\varphi<0$.  However for larger values of $\varphi_{c}$, we find that all models are physical, independent of the vector coupling parameter, $K$.  In section \ref{VFC} we show that $K\lesssim1$ based on considerations of the mass of neutron stars.  Therefore, from the demand for subluminal propagation of perturbations, we can conclude that $\varphi_{c}\gtrsim0.001$.  We note that this is a conservative estimate, and a true minimum value may be higher than this due to variations in the equation of state.  Moreover, if one can further constrain the value of $K$ to be smaller than unity, then the minimum cosmological value of the scalar field will also rise accordingly.  We also mention the caveat that we have assumed only a temporal component of the vector field which may alter these results.

\begin{figure}[h]
		\begin{center}
		\includegraphics[height=0.3\textwidth,width=0.4\textwidth]{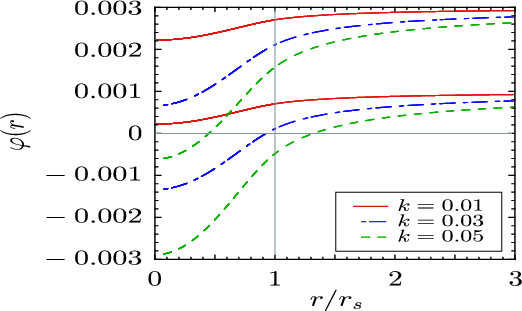}
		\caption{\label{varphi2} Scalar field as a function of radius with fixed $K=0.1$.  The top three lines have $\varphi_{c}=0.003$ and the bottom three lines are for $\varphi_c=0.001$.}  
		\end{center}
\end{figure}

Whilst in the above we assumed that $k=0.03$ according to \citet{bekenstein04}, it is worth exploring the variation of these values as a function of the scalar field coupling.  Figure \ref{varphi2} again shows the scalar field as a function of the radius, however this time we hold $K$ fixed at $0.1$ and vary the value of $k$.  One can see that when $\varphi_{c}=0.001$, only very small values of the scalar coupling permit models where $\varphi>0$ for all $r$.  However, when $\varphi_{c}=0.003$, models for $k\lesssim0.04$ are accepted.

\begin{figure}[h]
		\begin{center}
		\includegraphics[height=0.3\textwidth,width=0.4\textwidth]{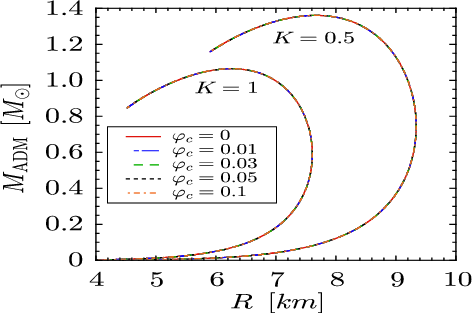}
		\caption{\label{varphiMR} ADM mass as a function of the physical radius, $R:=e^{-\varphi(r_s)}r_{s}$.  Here, $k=0.03$ and we show two different values of the vector coupling parameter, $K=0.5$ and $1.0$ for different cosmological values of the scalar field.}
		\end{center}
\end{figure}

An important quantity that will be considered throughout this article is the mass-radius relation, where we are using the observable ADM mass, $M_{\rm ADM}$, as a function of the physical radius, $R:=e^{-\varphi(r_s)}r_{s}$.  Figure \ref{varphiMR} shows the dependence of $\varphi_{c}$ on the structure of neutron stars for two different values of $K$.  One can see here that the dependence on the cosmological scalar field on the size of neutron stars is completely negligble.  This implies that, whilst $\varphi_{c}$ may evolve with cosmological time, the overall structure of neutron stars throughout this evolution will not significantly change.  

\subsection{Vector Field Coupling}\label{VFC}

The vector field coupling is possibly the least restricted parameter in the TeVeS theory.  To the best of our knowledge, the only discussion of this parameter is given by \citet{sanders06}, who postulated on phenomenological grounds that $K\lesssim10^{-7}$. In this article we will provide constraints on the vector field of $K\lesssim1$ based solely on observations of neutron stars.  It is difficult to further constrain the parameter using this method due to uncertainties in the equation of state.

\begin{figure}[h]
		\begin{center}
		\includegraphics[height=0.3\textwidth,width=0.4\textwidth]{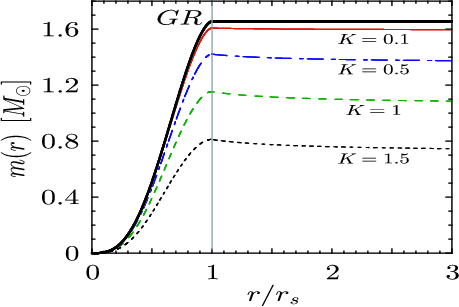}
		\caption{\label{massdep} Mass function, $m$, as a function of radius with $k=0.03$ and $\varphi_{c}=0.003$.  }  
		\end{center}
\end{figure}

\begin{figure}[h]
		\begin{center}
		\includegraphics[height=0.3\textwidth,width=0.4\textwidth]{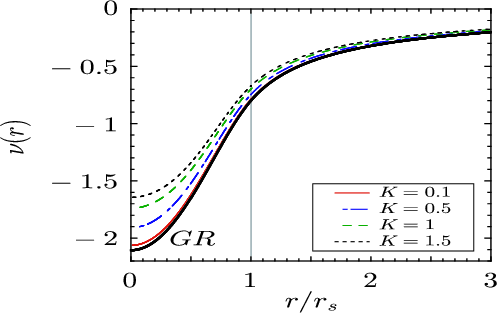}
		\caption{\label{nudep} Metric coefficient, $\nu$, as a function of radius with $k=0.03$ and $\varphi_{c}=0.003$.}  
		\end{center}
\end{figure}

Figures \ref{massdep} and \ref{nudep} show the structure of the mass function, $m$, and the metric coefficient, $\nu$, in the GR case as well as for different values of $K$, with $k=0.03$ and $\varphi_{c}=0.003$.  The stronger the vector coupling (i.e. higher $K$) implies the mass of the neutron star is decreasing.  We note that the mass function has a local maximum which is different from the asymptotic value as one would expect.  This is because the scalar field acts to decrease the overall mass of the star exterior to the stellar surface (this was also discussed briefly in \citet{giannios05}).  

\begin{figure}[h]
		\begin{center}
		\includegraphics[height=0.3\textwidth,width=0.4\textwidth]{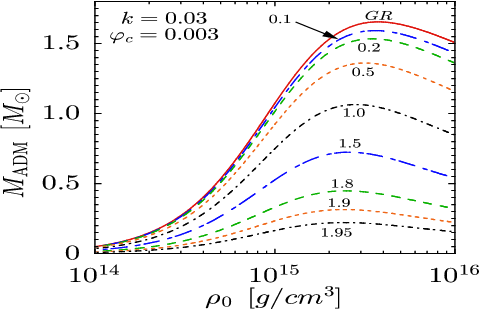}
		\caption{\label{rhocM} Central density, $\rho_{0}:=\rho(r=0)$ as a function of the ADM mass, $M_{\rm ADM}$ for varying values of the vector coupling constant $K$.  Here, the scalar coupling constant is held constant at $k=0.03$ and the cosmological value of the scalar field is $\varphi_{c}=0.003$.  Variations for the ADM mass are small for small values of the vector coupling constant $K$, however they grow as $K$ tends towards $2$.}  
		\end{center}
\end{figure}

\begin{figure}[]		
\begin{center}
		\includegraphics[height=0.3\textwidth,width=0.4\textwidth]{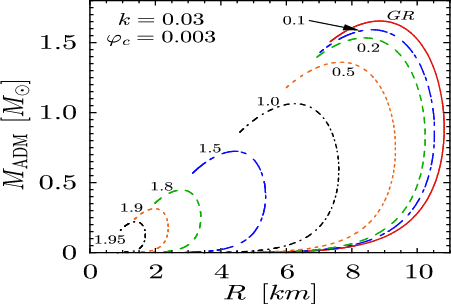}
		\caption{\label{RM} Mass-radius relation for various values of the vector coupling parameter $K$.  Again, $k=0.03$ and $\varphi_{c}=0.003$.  For small values of $K$ there is little difference in the mass-radius relation from the predictions from GR.  However there are significant differences for larger values of $K$, where the mass of a neutron star with the equation of state being considered here must be smaller than $M_{\rm ADM}\sim 1\,M_{\astrosun}.$} 
		\end{center}
\end{figure}

In figure \ref{rhocM} we show the dependence of the measurable ADM mass on the central density of the star, $\rho_{0}:=\rho(r=0)$, for varying values of the vector coupling parameter, $K$.  Figure \ref{RM} shows the ADM mass-radius relation of the star for the same models as in figure \ref{rhocM}.  In both these figures we use $k=0.03$ and $\varphi_{c}=0.003$.  One can see from these two figures that for small values of $K$, i.e. $K\lesssim0.05$, the deviations from GR are small, and indeed these effects are difficult to distinguish between variations around the equation of state and different values of $K$.  However, for larger values of $K\gtrsim0.05$, the deviations from GR begin to become considerable.  Indeed, in these models one can not create neutron stars with mass greater than approximately $1.3\,\, M_{\astrosun}$.  When the value of $K$ is greater than unity, neutron stars begin to look extremely different than those predicted in GR.  We find that, independent of the central density, one can not create a neutron star with mass greater than approximately one solar mass.  Moreover, the radius in these models is still less than approximately seven kilometres.  

\subsection{Scalar Field Coupling}\label{SFC}
The scalar field coupling has been constrained by \citet{bekenstein04} using planetary motions in the outer solar system to be $k\approx0.03$.  However, we note that this used a specific functional form of the free function $\F$, whereas in this paper our results have been shown to be independent of this function.  Therefore, in this section we explore the parameter space of $k$ in an attempt to place model independent constraints on its value.

\begin{figure}[h]
		\begin{center}
		\includegraphics[height=0.3\textwidth,width=0.4\textwidth]{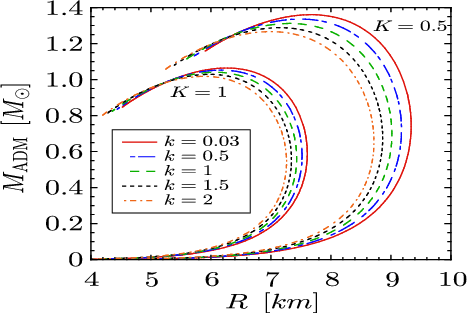}
		\caption{\label{kMR} Mass-radius relation for variations of the scalar field coupling parameter.  Here, $\varphi_{c} = 0.003$ and the vector coupling parameter is $K=0.5$ and $K=1.0$.}  
		\end{center}
\end{figure}

Figure \ref{kMR} shows the mass-radius relation for different values of the scalar field coupling parameter.  Here, $\varphi_{c}=0.003$ and $K=0.5$ and $1.0$.  One can see that that value of the scalar coupling parameter needs to be significantly altered to change the size of the neutron stars.  It is interesting to note that increasing the value of $k$ creates smaller neutron stars.  As mentioned, some models with larger values of $k$ may be unphysical due to the causality problem discussed in section \ref{CVSF}.  However, considering fig. \ref{varphiMR}, we can still conclude that the dependence of the scalar field coupling, $k$, on the structure of neutron stars is not so strong.

\section{Observational Consequences}\label{Obs}
Thus far, we have investigated the structure of neutron stars in TeVeS as distinct from GR.  It is worthwhile spending some time establishing how this can be seen observationally, and therefore provide independent verifications or falsifications of TeVeS.  We have seen in the above sections that the biggest variations in neutron star structure are due to the vector field coupling parameter, $K$.  Therefore, throughout the remainder of this section we choose to leave $k=0.03$ and $\varphi_{c}= 0.003$, and only vary the values of $K$.  It is worth noting at this point that if the discussion of \citet{sanders06} is correct (i.e. that $K<10^{-7}$), then the differences in neutron star predictions by GR and by TeVeS are observationally indistinguishable.  

Figures \ref{rhocM} and \ref{RM} suggest that the mass and radius of neutron stars in TeVeS are smaller than those in GR.  That is, a higher value of the vector coupling constant, $K$, implies smaller masses and radii for neutron stars.  The most accurate measurements of neutron star masses are from timing observations of binary pulsars, although significant progress is also being made using binaries with white dwarf companions (see \citet{lattimer07} and references therein).  Conservatively, one can say that neutron stars exist with masses greater than $\sim1.5 M_{\odot}$, although the true value may be significantly greater.  Taking this into account, together with the uncertainties in the equation of state that we have been using, suggests we can rule out the parameter space $K\gtrsim1$, although we note that this too is probably a conservative estimate.  Moreover, as TeVeS systematically predicts lower masses for a fixed equation of state, if a set of equations of state are excluded by the measurement of a high mass neutron star in GR, then these equations of state are also excluded in TeVeS.  For example, if the mass of a neutron star is measured at $M\gtrsim1.7 M_{\odot}$, then our adopted equation of state is ruled out in both GR and TeVeS.  

Another observable quantity is the redshift of atomic lines emanating from the surface of the neutron stars.  The redshift is given by
\begin{align}
	z=-1+\sqrt{\frac{\g_{tt,\infty}}{\g_{tt,s}}},\label{redshift}
\end{align}
where $\g_{tt,\infty}$ and $\g_{tt,s}$ are the values of the temporal component of the physical metric at radial infinity and the stellar surface respectively.  With this redshift factor $z$, the photon energy emitted from the stellar surface, $E_s$, can be connected to the energy measured at infinity, $E_\infty$, as $E_\infty=[1/(1+z)]E_s$ \cite{dedeo03}.  

\begin{figure}[h]
\begin{center}
		\includegraphics[height=0.3\textwidth,width=0.4\textwidth]{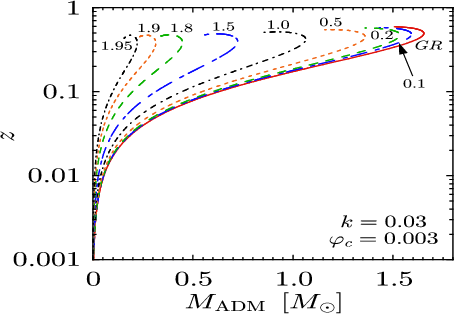}
		\caption{\label{Mz} Redshift as a function of the ADM mass for various values of $K$, with $k = 0.03$ and $\varphi_{c}=0.003$.}  
		\end{center}
\end{figure}

Figure \ref{Mz} shows the redshift at radial infinity of a photon emitted from the surface of the neutron star as a function of the mass for various values of $K$ with $k=0.03$ and $\varphi_{c}=0.003$.  Again, for small values of the vector coupling parameter the deviations from GR are small.  For larger values however, one finds a large redshift from extremely small values of the mass.  That is, for neutron stars of $M_{\rm ADM}\lesssim 1\,\, M_{\astrosun}$, one can still have redshift $z\sim 0.4$. 

\begin{figure}[h]
		\begin{center}
		\includegraphics[height=0.3\textwidth,width=0.4\textwidth]{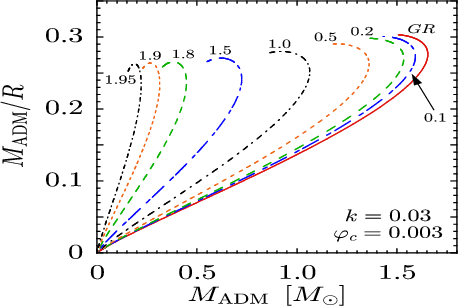}
		\caption{\label{MMonR} Compactness, $M_{\rm ADM}/R$, as a function of mass for various values of $K$, with $k=0.03$ and $\varphi_{c}=0.003$.}  
		\end{center}
\end{figure}

An exciting future probe of neutron stars and black holes is through gravitational wave astronomy.  Neutron stars in metric theories of gravity are known to exhibit pulsations of the spacetime, known as $w$-modes \cite{kokkotas92}.  These $w$-modes have frequencies, $f_{w}$, where $f_{w}R\propto M_{\rm ADM}/R$ (see \citet{kokkotas99} for a discussion in GR and \citet{sotani05} for scalar-tensor theory).  $M_{\rm ADM}/R$ is known as the {\it compactness}, and we plot this against the ADM mass for different values of the vector coupling, $K$, with $k=0.03$ and $\varphi_{c}=0.003$ in figure \ref{MMonR}.  If future gravitational wave detectors are able to reach the required sensitivity to accurately determine the compactness of neutron stars through $w$-mode oscillations, then these results coupled to the mass estimates from pulsar timing observations will be able to provide an independent constraint on the vector coupling constant, $K$.  Moreover, it has been shown that the relationship between the frequency of the $w$-mode and the compactness is almost independent of the equation of state \cite{kokkotas99,sotani05}.  This implies that gravitational waves will provide a unique probe of the appropriate theory of gravity which, unlike most other observations of neutron stars, does not depend on the equation of state.

\section{Conclusion}\label{Conc}
We have solved here the equations of hydrostatic equilibrium for a spherically symmetric, static distribution of a perfect fluid in Bekenstein's Tensor-Vector-Scalar (TeVeS) theory \cite{bekenstein04}.  Further imposing a specific polytropic equation of state has allowed us to analyse the differences in neutron star structure between those appearing in General Relativity and those in TeVeS.  We have shown that the form of neutron stars in TeVeS depends most strongly on the coupling parameter associated with the vector field, and used this to show that $K\lesssim1$.  Constraining this parameter any further is difficult with current observations given the uncertainties in the equation of state.  We have further shown that the cosmological value of the scalar field, $\varphi_{c}$, is constrained in this theory by demanding that causality not be violated in the interior of neutron stars.  In some ways, this is a counter-intuitive result, as the global, cosmological value is determined by local measurements.  This led to the conservative constraint that $\varphi_{c}\gtrsim0.003$, where the uncertainty is again due to the equation of state as well as the choice of vector field only having a temporal component.  

Using the results derived herein we have looked at observable quantities such as the mass-radius relation, surface redshift and compactness which may allow future, independent verifications or falsifications of the theory.  If General Relativity is the correct theory of gravity, then these observations are sufficient to determine the equation of state of neutron stars.  However, in alternative models of gravity there exists a degeneracy between the equation of state and the specific free parameters in the theory.  This implies that independent establishments of the neutron star equation of state are required in order to stringently constrain the free parameters of any theory.

This work has provided a first step in the study of neutron stars in TeVeS.  The first and most obvious extension is to allow for a non-zero radial component of the vector field.  To do this, one simply requires that the radial component vanish at spacelike infinity such that the theory does not pick out a preferred direction in space.  In this way, the vector field equation (\ref{vec}) will no longer be trivially satisfied.  Instead, this will provide one equation, together with the normalization criteria which will allow for the complete determination of the vector field in terms of the metric coefficients.  As was shown by \citet{giannios05}, the introduction of a non-zero radial component can considerably alter the structure and geometry of the black hole solutions, and we anticipate this to also be the case for neutron star solutions.  

With basic neutron star models in place, one can begin to study more complicated aspects of the observations in order to further constrain TeVeS.  For example, recent observations of quasi-periodic oscillations in the giant flares associated with soft gamma repeaters are believed to be associated with the global oscillations of strongly magnetized neutron stars (for example see \cite{israel05,strohmayer05,strohmayer06}).  Perturbations of neutron stars in General Relativity can successfully account for all of the frequencies associated with the quasi-periodic oscillations \cite{sotani08a}. Therefore, by performing a similar analysis on neutron stars in TeVeS one should theoretically be able to further constrain the various parameters associated with the theory.  This work is obviously non-trivial as one first needs to derive the coupled system of equations governing magnetic fields in TeVeS, at least to a first-order approximation.


\appendix
\renewcommand{\theequation}{A\arabic{equation}}
\setcounter{equation}{0}  
\renewcommand{\thesection}{\arabic{section}}
\setcounter{section}{0}
\section{Components of Einstein equations}\label{app}  
Here we show the various components of the modified Einstein field equations (\ref{Einstein}) which lead to the derivations of equations (\ref{Eins1}-\ref{Eins3}).  Firstly, the Einstein tensor, $G_{\mu\nu}$, which is derived with respect to the Einstein metric, (\ref{Emetric}), has four non-zero components
\begin{align}
	G_{tt}=&\frac{1}{r^{2}}e^{\nu-\zeta}\left(r\zeta'+e^{\zeta}-1\right),\\
	G_{rr}=&\frac{1}{r^{2}}\left(r\nu'-e^{\zeta}+1\right),\\
	G_{\theta\theta}=&\frac{r}{4}e^{-\zeta}\left(2\nu'-2\zeta'-\nu'\zeta'r+2\nu''+\nu'^{2}r\right).
\end{align}
The tensor $\Theta_{\mu\nu}$, as defined in equation (\ref{Theta}) is given in component form as
\begin{align}
	\Theta_{tt}=&\frac{K}{4}e^{\nu-\zeta}\left(\frac{\nu'^{2}}{2}-\nu'\zeta'+2\nu''+\frac{4\nu'}{r}\right)\notag\\
		&+8\pi G\left(1-e^{-4\varphi}\right)e^{\nu+2\varphi}\rhot,\\
	\Theta_{rr}=&-\frac{K}{8}\nu'^{2},\\
	\Theta_{\theta\theta}=&\frac{K}{8}e^{-\zeta}r^{2}\nu'^{2}=\frac{1}{\sin^{2}\theta}\Theta_{\phi\phi},
\end{align}
where all other components vanish.  The tensor $\tau_{\mu\nu}$, as defined in (\ref{tau}) is
\begin{align}
	\tau_{tt}=&\frac{e^{\nu-\zeta}}{2kG}\varphi'^{2},\\
	\tau_{rr}=&\frac{1}{2kG}\varphi'^{2},\\
	\tau_{\theta\theta}=&\frac{-r^{2}}{2kG}e^{-\zeta}\varphi'^{2}=\frac{1}{\sin^{2}\theta}\tau_{\phi\phi}.
\end{align}
Finally, defining $\tilde{\mathcal{T}}_{\mu\nu}:=\T_{\mu\nu}+\left(1-e^{-4\varphi}\right)\U^{\alpha}\T_{\alpha(\mu}\U_{\nu)}$, it is found
\begin{align}
	\tilde{\mathcal{T}}_{tt}=&\left(2e^{-2\varphi}-e^{2\varphi}\right)\rhot e^{\nu},\\
	\tilde{\mathcal{T}}_{rr}=&e^{-2\varphi}\Pt e^{\zeta},\\
	\tilde{\mathcal{T}}_{\theta\theta}=&e^{-2\varphi}r^{2}\Pt=\frac{1}{\sin^{2}\theta}\tilde{\mathcal{T}}_{\phi\phi}.
\end{align}

\acknowledgments{We thank the referee for fruitful suggestions regarding the manuscript.  PL and HS were supported via the Transregio 7 ``Gravitational Wave Astronomy'', financed by the Deutsche Forschungsgemeinschaft DFG (German Research Foundation).  PL was partially supported by the Australian Research Council's ``Commonwealth Cosmology Initiative'' (www.thecci.org: DP0665574).}

\bibliography{TOVTeVeS}
\end{document}